\documentclass[twocolumn,showpacs,preprintnumbers,amsmath,amssymb,superscriptaddress,groupedaddress,prx]{revtex4}

\pdfoutput=1
\usepackage{epsfig}                                                            
\usepackage{graphicx}
\usepackage{dcolumn}
\usepackage{color}
\usepackage{bm}
\usepackage{subfigure}

\begin{document}

\title{\bf Large magnetoelectric coupling in nanoscale BiFeO$_3$ from direct electrical measurements}

\author {Sudipta Goswami} \affiliation {Nanostructured Materials Division, CSIR-Central Glass and Ceramic Research Institute, Kolkata 700032, India}
\author {Dipten Bhattacharya} 
\email{dipten@cgcri.res.in} \affiliation {Nanostructured Materials Division, CSIR-Central Glass and Ceramic Research Institute, Kolkata 700032, India}
\author {Lynette Keeney} \affiliation {Tyndall National Institute, University College Cork, Lee Maltings, Dyke Parade, Cork, Ireland}
\author {Tuhin Maity} \affiliation {Micropower-Nanomagnetics Group, Microsystems Center, Tyndall National Institute, University College Cork, Lee Maltings, Dyke Parade, Cork, Ireland}
\author {S.D. Kaushik} \affiliation {UGC-DAE Consortium for Scientific Research, Bhabha Atomic Research Centre, Mumbai 400085, India}
\author {V. Siruguri} \affiliation {UGC-DAE Consortium for Scientific Research, Bhabha Atomic Research Centre, Mumbai 400085, India}
\author {Gopes C. Das} \affiliation {Department of Metallurgical and Materials Engineering, Jadavpur University, Kolkata 700032, India}
\author {Haifang Yang} \affiliation {Laboratory of Microfabrication, Institute of Physics, Beijing 100190, P.R. China}
\author {Wuxia Li} \affiliation {Laboratory of Microfabrication, Institute of Physics, Beijing 100190, P.R. China}
\author {Chang-zhi Gu} \affiliation {Laboratory of Microfabrication, Institute of Physics, Beijing 100190, P.R. China}
\author {M.E. Pemble} \affiliation {Tyndall National Institute, University College Cork, Lee Maltings, Dyke Parade, Cork, Ireland}
\author {Saibal Roy}
\email{saibal.roy@tyndall.ie} \affiliation {Micropower-Nanomagnetics Group, Microsystems Center, Tyndall National Institute, University College Cork, Lee Maltings, Dyke Parade, Cork, Ireland} 

\date{\today}

\begin{abstract}
We report the results of direct measurement of remanent hysteresis loops on  nanochains of BiFeO$_3$ at room temperature under zero and $\sim$20 kOe magnetic field. We noticed a suppression of remanent polarization by nearly $\sim$40\% under the magnetic field. The powder neutron diffraction data reveal significant ion displacements under a magnetic field which seems to be the origin of the suppression of polarization. The isolated nanoparticles, comprising the chains, exhibit evolution of ferroelectric domains under dc electric field and complete 180$^o$ switching in switching-spectroscopy piezoresponse force microscopy. They also exhibit stronger ferromagnetism with nearly an order of magnitude higher saturation magnetization than that of the bulk sample. These results show that the nanoscale BiFeO$_3$ exhibits coexistence of ferroelectric and ferromagnetic order and a strong magnetoelectric multiferroic coupling at room temperature comparable to what some of the type-II multiferroics show at a very low temperature.
\end{abstract}
\maketitle

\section{Introduction}
The coexistence of long-range ferro orders and a strong cross-coupling between the respective order parameters at nanoscale can open a floodgate of development of next generation low power spintronic devices where the spin structure is manipulated by electric field. The question of whether really such ferro orders and a strong cross-coupling of the order parameters persist at nanoscale assumes immense importance. Because of its room temperature multiferroicity, BiFeO$_3$ happens to be the most promising candidate for investigating the issue of multiferroicity at nanoscale. It has already been observed that the antiferromagnetic spin structure flops under a sweeping electric field \cite{Lebeugle} in a single crystal of BiFeO$_3$ at room temperature. Observation of such a strong coupling has cleared the age-old doubts \cite{Popov} about its multiferroicity in bulk form. In thicker films (of thickness $\sim$600 nm) too, switching of antiferromagnetic domains could be observed under an electric field \cite{Zhao}. Doubts, however, still remain about whether such a strong multiferroicity is retained in the nanoscale and, if so, down to what size limit can one expect to observe the persistence of ferroelectric order and multiferroic coupling. It has widely been reported that the magnetization improves in the nanoscale due to incomplete spin spiral, enhanced canting, and lattice strain \cite{Ren}. Ferroelectric order, on the other hand, is found to be retained in a film of thickness as small as $\sim$2 nm in experiments performed by one group \cite{Chu} but non-existent in particles of size below $\sim$18 nm in experiments performed by another \cite{Petkov}. From the piezoresponse force microscopy, it has earlier been shown that, in general, the size limit for ferroelectricity could be $\sim$20 nm \cite{Roelofs}. Even the experiments which revealed persistence of ferroelectric order in nanoscale BiFeO$_3$, had actually observed a mosaic texture of ferroelectric domains \cite{Chu}. More importantly, the main controversy seems to centre on whether the magnetoelectric coupling is strong in that size range. For future generation nanospintronic applications, nanoscale BiFeO$_3$ particles with stable single ferroelectric domains exhibiting coherent domain switching under a magnetic field or stable magnetic domains undergoing switching under an electric field need to be synthesized and investigated in detail. In this paper, we report the results of direct measurement of polarization versus electric field loops on the nanochains of BiFeO$_3$ under zero and $\sim$20 kOe magnetic field at room temperature. We found that the remanent polarization (P$_r$) is suppressed by nearly 40\% under the magnetic field. Rietveld refinement of the high resolution powder neutron diffraction data showed that the structural noncentrosymmetry increases as the particle size is brought down from bulk to a scale of $\sim$30 nm. The piezoresponse force microscopy (PFM), on other hand, revealed the isolated single crystalline nanoparticles to be monodomain in nature containing single ferroelectric domains undergoing 180$^o$ phase evolution under a dc electric field. The particles also depict nearly an order of magnitude higher saturation magnetization than that for the bulk sample at room temperature. These results show that the nanochains of BiFeO$_3$, comprising of particles with switchable single ferroelectric domains and large saturation magnetization, exhibit a strong magnetoelectric coupling at room temperature which can be utilized in novel nanospintronic applications.

\begin{figure}[!ht]
  \begin{center}
    \includegraphics[scale=0.20]{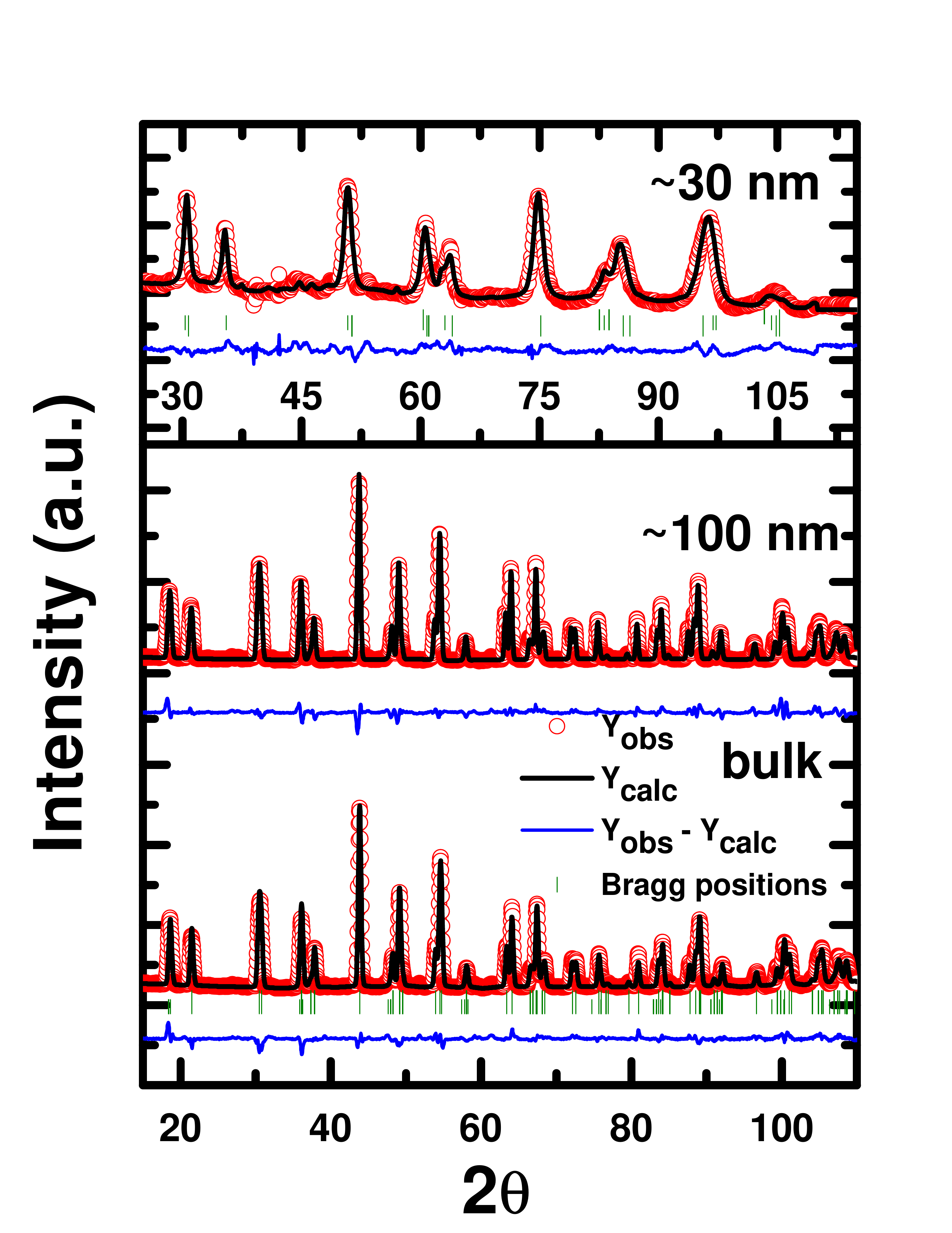} 
    \end{center}
  \caption{(color online) The powder neutron diffraction data and their refinement by FullProf for samples of different particle sizes; while the wavelength corresponding to the topmost pattern is 2.417 \AA, it is 1.480 \AA  for the other patterns. }
\end{figure}

\begin{figure}[!ht]
  \begin{center}
    \includegraphics[scale=0.35]{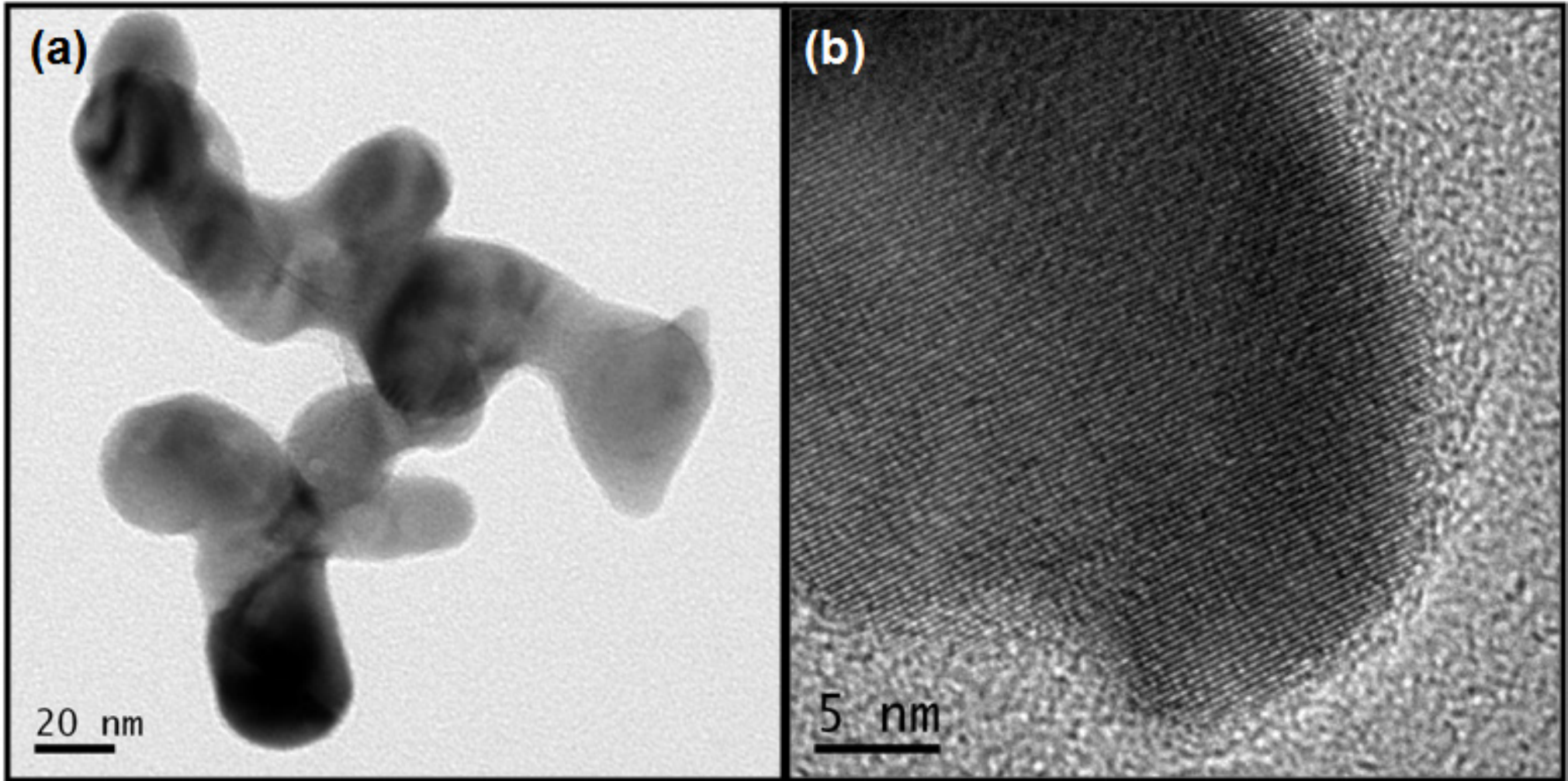} 
    \end{center}
  \caption{The (a) transmission electron microscopy image and (b) the high-resolution TEM image for the nanoparticles. The nanoparticles turn out to be single crystalline with (110) plane (d = 2.762 \AA) oriented perpendicular to the beam direction. }
\end{figure}

\begin{figure*}[!ht]
  \begin{center}
    \includegraphics[scale=0.60]{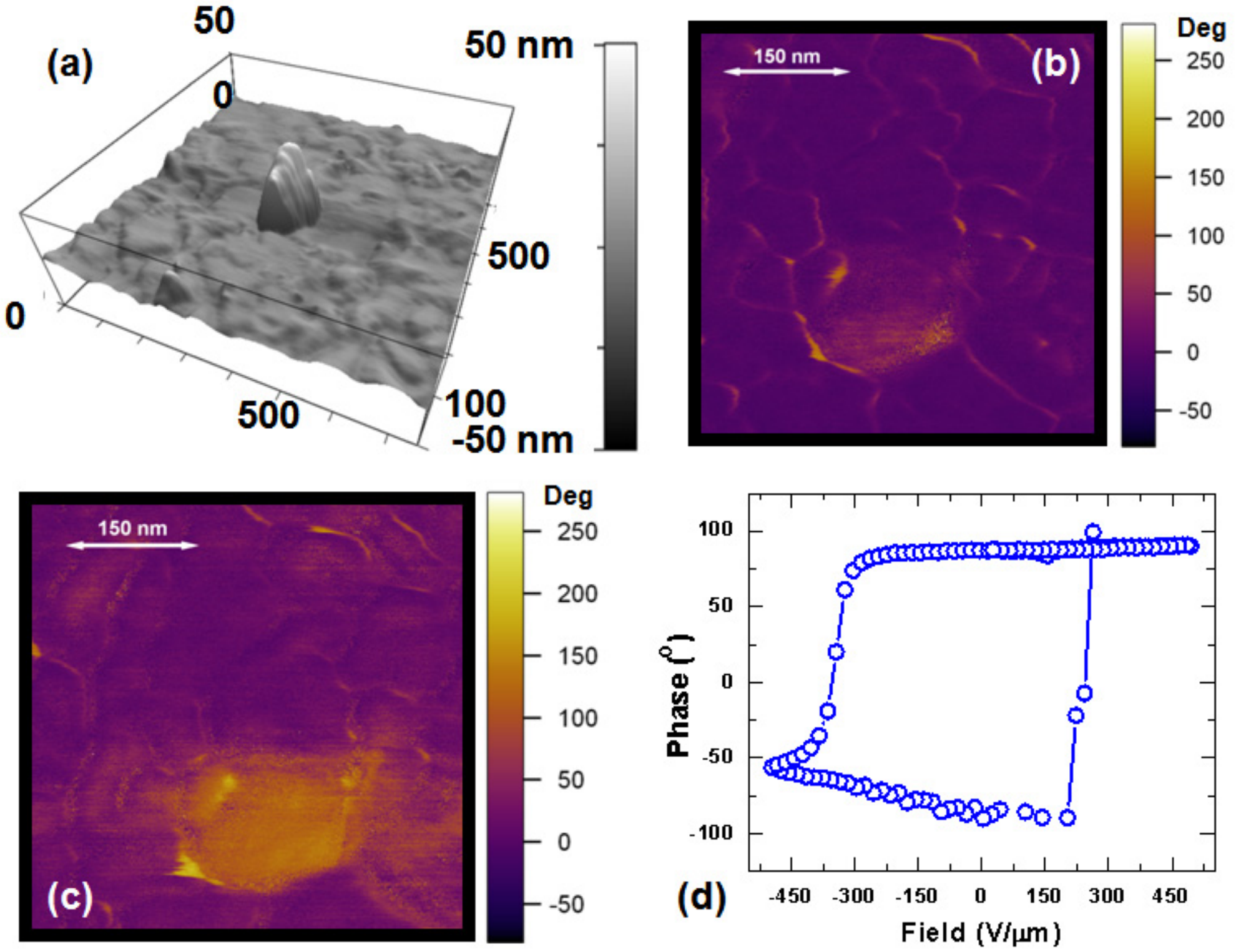} 
    \end{center}
  \caption{(color online) The (a) AFM and phase contrast PFM images under (b) zero field and (c) poling under 10 V of an isolated nanoparticle of BiFeO$_3$ in vertical scan; poling leads to a change in color from purple to yellow because of $\sim$90$^o$ switching of polar domain; the polarization is oriented vertically after poling; (d) the complete phase evolution hysteresis loops from switching spectroscopy PFM measurement. }
\end{figure*}

\section{Experiments}
The nanoparticles of BiFeO$_3$ were prepared by sonochemical synthesis route \cite{Goswami-1} while the bulk sample was prepared by solid-state reaction process. The high resolution powder neutron diffraction patterns were recorded at the multi-position-sensitive detector based focusing crystal diffractometer set up by UGC-DAE-CSR Mumbai Centre at the National Facility for Neutron Beam Research (NFNBR), Dhruva Reactor, Mumbai and also at the D20 diffractometer of the Institute Laue-Langevin, Grenoble, France. The data has been refined by FullProf. Piezoresponse force microscopy (PFM) was carried out using an Asylum Research MFP-3D$^{TM}$ atomic force microscope in contact mode, equipped with a HVA220 Amplifier. The vertical and lateral single frequency (drive frequency of 20 kHz) and vertical dual resonance tracking PFM (DART-PFM) modes were employed. The DART-PFM method uses the cantilever resonance frequency to boost the piezo signal in the vertical direction, while reducing the crosstalk between changes in sample-tip contact stiffness and the PFM signal by tracking the resonance frequency based on amplitude detection feedback. The isolated nanoparticles of BiFeO$_3$ were dispersed in isopropyl alcohol and deposited on a Si/SiO$_2$ substrate coated with gold. The gold coating serves as the bottom electrode. Olympus AC240TM Electrilevers and Pt coated silicon cantilevers (Al reflex coated, 70 kHz resonant frequency, 320 kHz contact resonance frequency) of tip radius 28$\pm$10 nm were used for PFM imaging. The amplitude and the phase contrast images were captured before and after poling by vertical application of an external field (10 V) by PFM probe. Vertical hysteresis loop measurements were carried out using a triangular shape waveform comprised of pulse DC bias voltage (10 V) and AC signal (5.5 V) in order to record the evolution of the domains by switching spectroscopy PFM (SS-PFM). The waveform was cycled twice at a frequency of 0.2 Hz with 100 AC steps per waveform. In order to avoid electrostatic contributions to the signal, the piezoresponse hysteresis loops were recorded in the off-field state. The magnetic measurements were carried out in a SQUID magnetometer (MPMS XL, Quantum Design) across 5-300 K under a maximum field H$_m$ of 50 kOe. The nanoparticles, deposited on Si/SiO$_2$ substrate, were also patterned by e-beam lithography (EBL) and focused ion beam milling/deposition (FIB) with gold and tungsten electrodes, respectively, for direct measurement of the P-E (polarization versus electric field) loops. The Precision LC loop tracer of Radiant Inc., USA was used for the measurements. 

\section{Results and Discussion}
In Fig. 1, we show the room temperature powder neutron diffraction data and their refinement. The space group appears to be R3c with rhombohedral Bravais lattices in all the cases. The polarization axes are along the $<$111$>$ direction for such a structure giving rise to eight possible directions of polarization. The fit statistics and the structural parameters determined from the refinement have been given in the supplementary document \cite{supplementary}. Interestingly, the structural noncentrosymmetry appears to have improved as the particle size is brought down from bulk to a scale of $\sim$30 nm - it increases from $\sim$1.48 \AA in bulk to $\sim$1.67 \AA in $\sim$100 nm particles to, finally, $\sim$1.73 \AA in particles of size $\sim$30 nm. This could result from enhanced lattice strain of the finer particles \cite{Goswami-1}. The strain has increased from 0.015\% to 0.1\% as the particle size reduces from bulk to nanoscale. The propagation vector for the magnetic phase is found to be k = (0, 0, 0) for all the cases. The magnetic moments turn out to be 3.24(5), 3.50(3), and 3.98(6) $\mu_B$/Fe for the bulk, $\sim$100 nm, and $\sim$30 nm particles, respectively. In Fig. 2, we show the representative transmission electron microscopic (TEM) and the high-resolution TEM (HRTEM) images of the nanoparticles. The individual nano-sized particles are single crystalline with (110) planes oriented perpendicular to the direction of the beam. These single crystalline particles are expected to contain single ferroelectric domains. Indeed, from the PFM measurements, it appears that the isolated nanoparticles of even bigger sizes are monodomain in character. In Fig. 3a, we show a representative topography image of an isolated nanoparticle ($\sim$150 nm$\times$$\sim$150 nm$\times$$\sim$50 nm) of BiFeO$_3$ on which the PFM was carried out. Fig. 3b shows the phase contrast image of the particle, captured in the vertical dual resonance tracking PFM (DART-PFM) mode, under zero DC electric field. PFM images of the unpoled BiFeO$_3$ particles illustrate relatively weak amplitude contrast and a phase contrast of $\sim$15$^o$. The line scan across the phase contrast image, shown in the supplementary document \cite{supplementary}, maps the phase evolution angle and the particle-background contrast in the unpoled state. The lack of even stronger electromechanical response and consequent amplitude contrast, however, could result from surface deformation generated by tip-surface electrostatic forces rather than inherent electromechanical coupling in the self-poled state. Given the relatively weak signal to background noise ratio for the amplitude contrast image, it is not shown here. Fig. 3c shows the phase contrast image after poling by a vertical DC bias of 10 V. A phase contrast of $\sim$90$^o$ between the BiFeO$_3$ particles and the background is observed after poling and the color of the domain changes from purple to yellow as a result of emergence of vertical polar domains. The domain switching angles obtained from line scanning on the PFM images have been given in the supplementary document \cite{supplementary}. Since the nanoparticles of BiFeO$_3$ retain the rhombohedral symmetry (space group R3c) with polarization oriented along the $<$111$>$ direction (in pseudocubic notation), the polarization could either rotate by $\sim$71$^o$ or $\sim$109$^o$, or switch by $\sim$180$^o$ \cite{Ramesh}. The complete $\sim$180$^o$ switching in switching spectroscopy PFM (SS-PFM) has been observed, in the present case, under a reversible DC bias (Fig. 3d). The domain switching under an external DC bias is a signature characteristic of switchable ferroelectricity. The phase evolution hysteresis loop, of course, depicts a small vertical offset originating from the difference in charge conduction under different polarities between top (Pt) and bottom (Au) electrodes \cite{Zheng} and/or presence of a polar nonferroelectric and hence nonswitchable region at the surface. Particles with multi-domains, on the other hand, depict a sizable vertical shift of the hysteresis loop resulting from non-switchable frozen part \cite{Rodriguez}. The horizontal asymmetry of the loop results from self-biasing within the particles. The shape of the phase loop demonstrates incomplete saturation of the hysteresis, an indication of the presence of domains that are still evolving. This type of saturation is usual in weak ferroelectrics as much higher fields are typically required to switch the domain polarization. The lateral PFM scans, measured using the single frequency PFM mode, did not demonstrate significant phase contrast or amplitude piezoresponse above the signal to noise level. In order to avoid topography cross-talk in the single frequency mode, imaging must be performed far from the contact resonance. As there is no resonance enhancement in the single frequency PFM mode, the relatively weak piezo signal from the BiFeO$_3$ is not boosted and reliable images could not be obtained. Nonetheless, the vertical scan phase images on poling imply that the ferroelectric order does prevail and particles as big as $\sim$150 $\times$ $\sim$50 nm are, in fact, even single domain in nature. No evidence of the presence of domain texture could be observed within the resolution limit of the PFM scan. This is an important observation as the ferroelectric polarization as well as multiferroic coupling is expected to be maximum in a single-domain system. A single domain particle rotates coherently under poling whereas the multi-domain particles undergo complicated switching dynamics with formation of nucleation centres and their growth under poling \cite{Rodriguez}. The yellowish color at the edges of the particle in Fig. 3b is likely due to imaging artifact.

\begin{figure}[!ht]
  \begin{center}
    \includegraphics[scale=0.30]{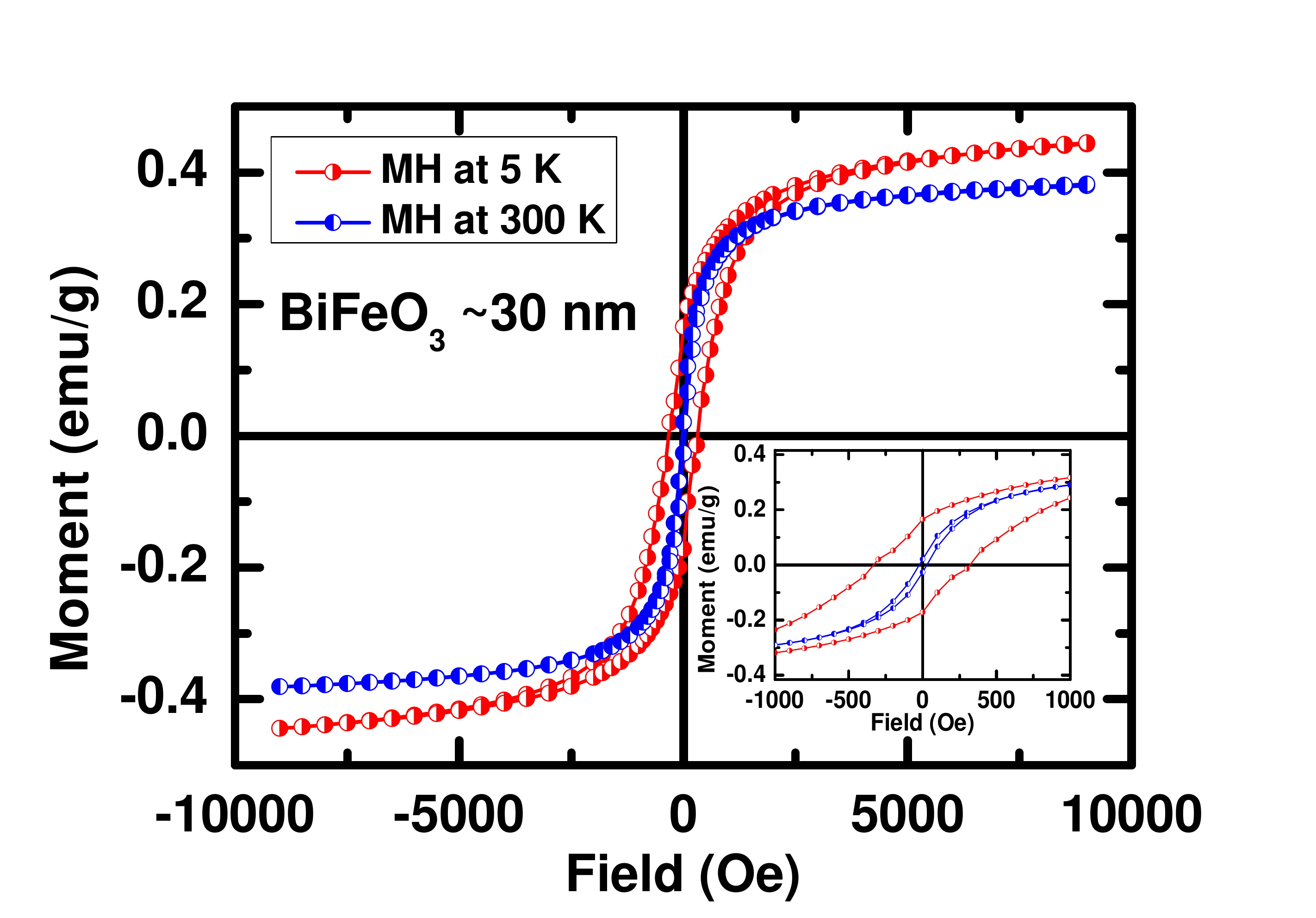} 
    \end{center}
  \caption{(color online) The magnetic hysteresis loop showing the soft ferromagnetic behavior of nanoscale BiFeO$_3$; the portion of the loop near origin is blown up in the inset to demonstrate absence of any exchange bias whatsoever thus proving the sample to be free from any magnetic impurity. }
\end{figure}

\begin{figure}[!h]
  \begin{center}
   \subfigure[]{\includegraphics[scale=0.80]{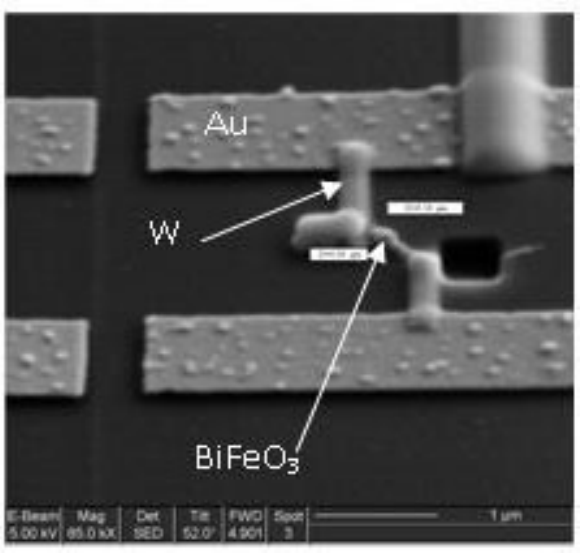}}
    \subfigure[]{\includegraphics[scale=0.25]{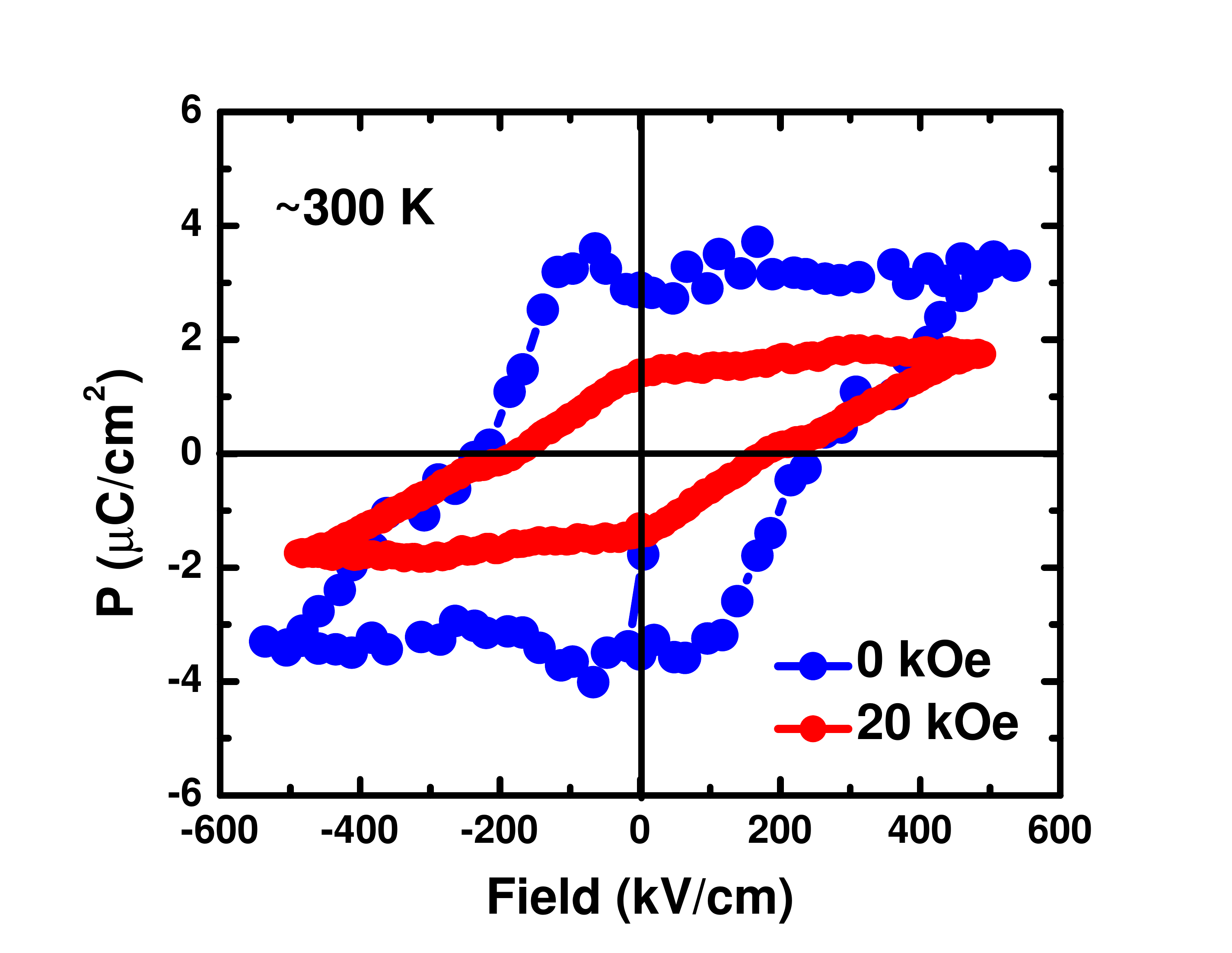}} 
    \end{center}
  \caption{(color online) The (a) scanning electron microscopy image of the patterned nanochain of BiFeO$_3$ and (b) the remanent hysteresis loop measured at room temperature under zero and $\sim$20 kOe magnetic field. }
\end{figure}

In Fig. 4, we show the magnetic hysteresis loops measured at 5 K and room temperature. This result demonstrates that the nanoscale BiFeO$_3$ of average particle size $\sim$30 nm are ferromagnetic at room temperature with nearly an order of magnitude higher saturation magnetization ($\sim$0.0244 $\mu_B$/Fe) and coercivity ($\sim$50 Oe). The measurements have been carried out on powdered sample of mass $\sim$0.36 g. The absolute saturation magnetic moment turns out to be $\sim$0.144 emu. Ferromagnetism with large magnetization has already been observed in nanosized particles of BiFeO$_3$, as a result of imcomplete spiral and enhanced canting of the spins, by others \cite{Park}. This sample is free from any secondary impurity phase such as Bi$_2$Fe$_4$O$_9$. In our earlier work \cite{Maity-1,Maity-2}, we showed that sizable exchange bias (both spontaneous and conventional) could be observed in systems where Bi$_2$Fe$_4$O$_9$ is present by $\sim$6 vol\%. In the present case, no exchange bias could be observed (inset of Fig. 4) which indicates that the presence of any secondary phase here is negligible. The ferromagnetism is intrinsic and emerges in nanoscale because of incomplete spin spiral (spiral wavelength is $\sim$62 nm), enhanced canting angle, and increased lattice strain. Therefore, the results presented in Figs. 3 and 4 clearly show coexistence of multiferro orders - ferroelectric and ferromagnetic - in nanoscale BiFeO$_3$. 

In Fig. 5a, we show a scanning electron microscopy image of a typical nanochain of BiFeO$_3$ patterned by e-beam lithography (EBL) and focused ion beam milling/deposition (FIB). Direct measurement of the remanent hysteresis loop was carried out at room temperature on this sample both under zero and $\sim$20 kOe magnetic field. In Fig. 5b, we show the loops. The remanent hysteresis loops were measured by Sawyer-Tower triangular electric field profile following a certain protocol in order to ensure elimination of all the contribution from resistive leakage and nonremanent polarization. This is necessary here as the ferroelectric polarization is small and the sample contains disorder at the interfaces between the particles in a chain. In this protocol, two hysteresis loops are measured following two different logics - logic1 and logic0. The first hysteresis loop is constituted using contributions from both remanent (i.e., switchable) and nonremanent (i.e., nonswitchable) polarization while the second one is constituted of non-remanent polarization alone. Subtraction of the latter loop from the former one yields the remanent hysteresis loop \cite{Evans,Wang}. As discussed below, both the logic1 and logic0 protocols use two prepolarization (preset) and two measurement voltage pulses to construct two hysteresis loops. No mesurement is done during the prepolarization pulses while data are actually measured during the measurement pulses. The sequence of events is as follows. In the case of logic1, initially, a preset pulse is used to prepolarize the ferroelectric domains of the sample along a certain direction. During this period no measurement is carried out. Next, another pulse with polarity opposite to that of the preset pulse is used. This pulse, therefore, switches the ferroelectric polarization from its initial direction to the direction of this second pulse. The corresponding polarization current is measured during this pulse and the polarization is estimated from integration of the current profile. This measurement, therefore, includes the contributions from both switchable and nonswitchable polarization as the nonswitchable polarization always contributes. Even though the entire measurement pulse comprises of one positive and one negative voltage pulse, actual measurement is carried out only during that voltage pulse whose polarity is opposite to the polarity of the preset pulse. From this measurement one half of the polarization versus electric field loop is constructed. Once this half of the loop is constructed, another preset and measurement pulses are used. This pulse train has opposite polarity to that of the initial train. Therefore, from this measurement, opposite half of the loop could be obtained. Once a complete loop is constructed in logic1, next set of voltage pulses (prepolarization and measurement) are used in logic0 in order to measure the nonswitchable polarization alone. In this case of logic0, polarity of both the preset and measurement pulses are kept the same in order to record the contributions from the nonremanent (i.e., nonswitchable) polarization. The remanent polarization has already been switched. Subtraction of the loop obtained in logic0 from the one obtained in logic1 yields the remanent hysteresis loop which gives an accurate measure of the intrinsic hysteretic ferroelectric polarization of the sample. The supplementary document \cite{supplementary} gives the actual voltage pulses used in logic1 and logic0 for the measurements as well as the loops obtained following them. The remanent hysteresis loops (Fig. 5b) obtained from this measurement protocol prove the persistence of switchable ferroelectric domains in $\sim$30 nm particles of BiFeO$_3$. The loops measured following standard protocol turn out to be characteristic of lossy dielectric and do not offer intrinsic hysteretic polarization because of small polarization and large loss. It is, of course, possible to arrive at the intrinsic remanent polarization by subtracting the contribution of the polarization due to leakage current to the overall polarization obtained in the standard measurement protocol. The complete leakage current versus field profile can be used to calculate the polarization due to leakage using the framework proposed by Fina $\textit{et al}$ \cite{Fina}. We also point out that in the remanent hysteresis loops (Fig. 5b) not much vertical or horizontal displacement with respect to the origin could be noticed. This is in contrast to the observation made in PFM and possibly results from identical electrode material (tungsten) at both the ends of the sample (Fig. 5a). In the case of the PFM, the top (Pt) and the bottom (Au) electrodes differ. The phase evolution hysteresis loop (Fig. 3d) obtained from SS-PFM and the remanent hysteresis loops (Fig. 5b) obtained from direct electrical measurements by Sawyer-Tower circuit, therefore, conclusively prove the persistence of long-range ferroelectric order in nanoscale BiFeO$_3$. We now turn our attention to the remanent hysteresis loop obtained under a magnetic field. It is noteworthy that the remanent polarization is suppressed along with a decrease in the loop area under $\sim$20 kOe magnetic field (Fig. 5b). The polarization drops by $\sim$40\% [= $\frac{P_r(0)-P_r(20 kOe)}{P_r(0)}$ $\times$100]. It appears, then, that a substantial change in the polarization occurs under the application of the magnetic field. The magnitude of the polarization across the entire nanochain, of course, is found to be small in comparison to that expected based on the structural noncentrosymmetry data for a unit cell and the observation in PFM that isolated nanoparticles are monodomain in nature. This could possibly be due to poor particle to particle connection which introduces surface effects. In fact, study of the dielectric spectroscopy on such a nanochain revealed presence of separate relaxation spectra for the interface and the bulk \cite{Goswami-2}. The suppression of polarization under a magnetic field observed in this direct electrical measurement on the nanochain is $\textit{consistent}$ with the results obtained from the neutron diffraction experiments \cite{Goswami-3,Goswami-4}. The plot of noncentrosymmetry as a function of temperature exhibits a suppression around the magnetic transition point T$_N$ \cite{Goswami-4}. Application of 50 kOe magnetic field at room temperature (i.e., at well below the T$_N$) also suppresses the noncentrosymmetry \cite{Goswami-3}. From the direct electrical measurements, the suppression of polarization under 20 kOe field is found to be $\sim$40\%. This result clearly proves that significant magnetoelectric coupling prevails in particles as fine as $\sim$30 nm. Such a clear proof of the prevalence of magnetoelectric coupling in the nanoscale still does not exist in the literature on BiFeO$_3$. It is important to mention here that the extent of change in ferroelectric polarization observed in nanoscale BiFeO$_3$ is comparable to the change ($\sim$30-90\%) observed in the ferroelectric polarization under a magnetic field ($\sim$20-90 kOe) for well-known type-II multiferroics such as TbMnO$_3$ \cite{Staruch,Kimura}, DyMnO$_3$,\cite{Lu} CuCrO$_2$ \cite{Seki} etc. in bulk, thin film, or single crystal form with 3D or 2D magnetic structure. However, in these compounds such a strong multiferroicity is observed only at a very low temperature ($\le$10-30 K). 
        
It has been argued by Lebeugle $\textit{et al}$. \cite{Lebeugle} that in bulk BiFeO$_3$ the spiral magnetic structure gives rise to a ferroelectric polarization via inverse Dzialoshinskii-Moriya exchange coupling interaction \cite{Mostovoy}. This, in turn, couples to the ferroelectricity originating from Bi-O bond covalency and yields a multiferroic coupling. In the nanoscale, the spiral is incomplete and cannot yield any polarization. However, magnetization (M) is large by one order of magnitude because of incomplete spiral and enhanced canting. This is expected to increase the linear $\vec{\bf{M}}.\vec{\bf{E}}$ (magnetoelectric) coupling. Structurally also, enhanced magnetization is associated with enhanced antiferrodistortive rotation of FeO$_6$ octahedra around the polarization axis [111] which, in turn, is expected to influence the off-center distortion of the cations and anions and hence the polarization \cite{Spaldin}. It has been shown \cite{Spaldin} from the symmetry arguments that the rhombohedral R3c structure is arrived from ideal cubic Pm$\bar{3}$m by inducing polar distortion along the [111] axis and antiferrodistortive rotation of FeO$_6$ octahedra around [111]. The magnetization, contained within the (111) plane and associated with the rotation of FeO$_6$ octahedra, if changed, leads to a rotation or even reversal of the polarization to a particular symmetry-allowed orientation state. Under a magnetic field, then, enhanced magnetization could lead to a reorientation or reversal of the polarization which, if averaged over an ensemble of nanoparticles, could give rise to a suppression of the net polarization. More microscopically, it occurs because of ionic displacement under a magnetic field via inverse Dzialoshinskii-Moriya exchange coupling interaction. In order to estimate the extent of ionic displacement under a magnetic field, we analyze the results obtained from Rietveld refinement of the room temperature neutron diffraction data \cite{Goswami-3} recorded under zero and 50 kOe magnetic field for nanoscale BiFeO$_3$. The ionic positions for Fe and O ions in a unit cell were refined while the position of the Bi ion was kept fixed. It appears that the Fe and O ions are displaced by $\sim$0.005 \AA/T and $\sim$0.02 \AA/T, respectively. The magnetization is enhanced by nearly 50\%/T and the lattice strain is suppressed by nearly 10\%/T at room temperature. It has earlier been shown that application of a magnetic field and consequent rise in magnetization gives rise to a femtoscale ionic displacement in a type-II multiferroic TbMnO$_3$ system because of antisymmetric Dzialoshinskii-Moriya exchange interaction \cite{Walker}. This leads to a substantial change in the polarization and has turned out to be the origin of the strong multiferroic coupling. Incidentally, the ionic displacement observed here in nanoscale BiFeO$_3$, under a magnetic field, appears to be more than an order of magnitude higher than what has been observed in TbMnO$_3$. It is also important to mention that strong magnetoelectric multiferroic coupling could be observed in type-II multiferroics, such as TbMnO$_3$, DyMnO$_3$, TbMn$_2$O$_5$ etc. \cite{Goto}, as magnetic structure itself yields the ferroelectric polarization. In the present case of nanoscale BiFeO$_3$, large magnetoelectric coupling is observed because of large magnetization and stronger ferromagnetic component arising out of incomplete spiral and enhanced canting even though ferroelectric order has a different origin. We further mention that even though the magnetoelectric effect observed here is primarily intrinsic in nature, the interface can have a finite contribution to the overall effect in the way described below. The disordered interface itself is expected to exhibit magnetoelectric effect since such a disordered spin and polar glass system is known to exhibit magnetoelectric effect \cite{Shvartsman,Choudhury}. The nonswitchable polarization component of the interface (i.e., dead layer), in fact, gives rise to the depolarizing field which reduces the intrinsic ferroelectric polarization \cite{Stengel,Bratkovsky,Alpay}. Because of finite magnetoelectric effect exhibited by this nonswitchable polarization of the interface, the depolarizing field itself is expected to change under a magnetic field. This changed depolarizing field, in turn, is expected to give rise to a changed influence on the intrinsic polarization. This change in intrinsic polarization as a result of changed depolarizing field is in addition to the intrinsic size effect related change in the ferroelectric polarization. Therefore, it seems that apart from significant intrinsic change in the ferroelectric polarization under a magnetic field, a small change can also result from change in the depolarizing field under a magnetic field. Though comparison of the extent of suppression of ferroelectric polarization under a magnetic field, noticed in neutron diffraction \cite{Goswami-3} and direct electrical measurements is hiniting at a possible influence of magnetoelectric effect of the disordered interfaces in this direct electrical measurement, a quantitative estimation of the contribution of the magnetoelectric effect at the interface on the intrinsic effect of the core of the particles requires controlled introduction of the disorder within an ordered region.    

\section{Summary}
In summary, we offer a conclusive evidence of the presence of ferroelectric order and strong magnetoelectric coupling in nanoscale BiFeO$_3$ from the piezoresponse force microscopy on isolated nanoparticles and direct electrical measurements on nanochains patterned by electron beam lithography and focused ion beam milling/deposition. The remanent polarization, in a chain of $\sim$30 nm particles, is found to have been suppressed by nearly 40\% under a magnetic field of $\sim$20 kOe at room temperature. Such a strong suppression of the remanent polarization is induced by large scale ionic displacement under a magnetic field via inverse Dzialoshinskii-Moriya exchange interaction. This is comparable to what has been observed in even those multiferroics which are well known for very strong coupling. More importantly, this has been observed at room temperature. The discovery of such a strong magnetoelectric coupling in nanoscale BiFeO$_3$ heralds a new beginning in the area of nano-spintronics with nano-multiferroics at its core. 

\begin{center}
$\textbf{ACKNOWLEDGMENTS}$
\end{center}

This work has been supported by Indo-Ireland joint program (DST/INT/IRE/P-15/11), ISCA grant (SFI: 12/ISCA/2493), Science Foundation Ireland (SFI) Principal Investigator (PI) Project No. 11/PI/1201 and SFI FORME Strategic Research Cluster Award number 07/SRC/I1172. One of the authors (S.G.) acknowledges support from a Research Associateship of CSIR.

\end{document}